\documentclass[%
 reprint,
 amsmath,amssymb,
 aps,
]{revtex4-2}
\usepackage{graphicx, epstopdf}% Include figure files
\usepackage{subcaption}
\usepackage{comment}
\usepackage{dcolumn}% Align table columns on decimal point
\usepackage{bm}% bold math
\usepackage[utf8]{inputenc}
\usepackage[T1]{fontenc}
\usepackage{mathptmx}
\usepackage[wby]{callouts}
\usepackage{url}
\usepackage {hyperref}
\usepackage{overpic}
\begin{document}
\preprint{AIP/123-QED}
\title{Retrieving Quality Factors for Reflection \& Transmission Measurements }

\author{Juliang Li}
\email{juliang.li@anl.gov}
\author{P. Barry, C. Change}
\affiliation{ 
High Energy Physics Division, Argonne National Laboratory%\\This line break forced with \textbackslash\textbackslash
}%
% \homepage{http://www.Second.institution.edu/~Charlie.Author.}

\begin{abstract}
This article presents pedagogical explanation of retrieving the resonance parameters $Q_{L}$, $Q_{o}$ and $Q_{c}$ from both reflection and transmission measurement of microwave resonator. Here $Q_{L}$ stands for the total or loaded quality factor (Q), $Q_{o}$ is the internal Q and $Q_{c}$ is the coupling or external Q. Matlab Code based on the methods is available for download for direct calculation of the Qs.\cite{lighq}
\end{abstract}
\maketitle
\section{model of reflection measurement}
\begin{figure}[h!]
% figure source: 'Trans and Ref Resonators.ppt',  juliang.li@anl.gov drive
   \begin{subfigure}[t]{0.22\textwidth}
%raisebox{-\height}{\includegraphics[width=\textwidth]{cirRef.png}}
   \begin{overpic}[abs,unit=1pt,scale=.28]{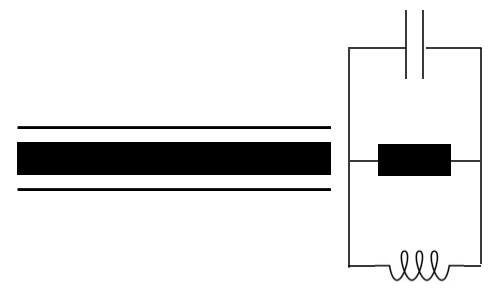}
   \put(3,40){\color{black}{(a)}}
   \end{overpic}
   \end{subfigure}
   \hfill
   \begin{subfigure}[t]{0.22\textwidth}
%   \raisebox{-\height}{\includegraphics[width=\textwidth]{cirRefPara.png}}
\begin{overpic}[abs,unit=1pt,scale=.25]{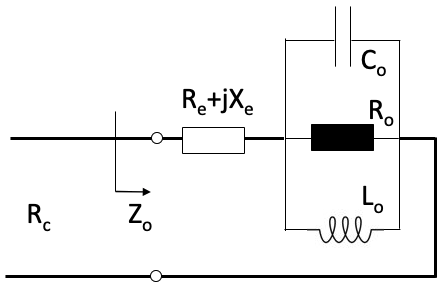}
%\put(92,80){\color{black}{(b)}}
\end{overpic}
   \end{subfigure}
      \begin{subfigure}[t]{0.22\textwidth}
%   \raisebox{-\height}{\includegraphics[width=\textwidth]{cirTran.PNG}}    
   \begin{overpic}[abs,unit=1pt,scale=.25]{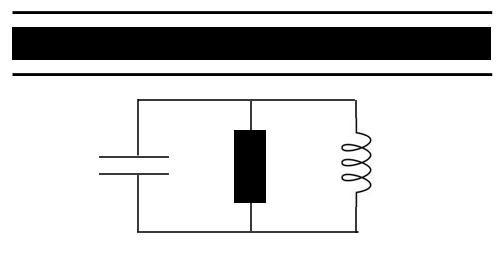}
   \put(3,4){\color{black}{(b)}}
   \end{overpic}
   \end{subfigure}
   \hfill
   \begin{subfigure}[t]{0.22\textwidth}
%   \raisebox{-\height}{\includegraphics[width=\textwidth]{cirTranPara.png}}
   \begin{overpic}[abs,unit=1pt,scale=.25]{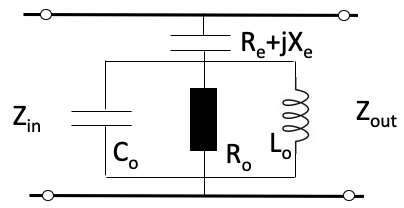}
%\put(92,80){\color{black}{(b)}}
\end{overpic}
   \end{subfigure}
\caption{\footnotesize{schematic of two typical resonator measurement circuits: reflective (\textbf{a}) and transmission (\textbf{b}). The right figures are the corresponding equivalent circuits with definition of the electronic components.}}
\label{cTranRef}
\end{figure}

For reflective type resonator as shown in figure \ref{cTranRef} (a) the transmitted signal measured by the network analyzer is given as: \cite{QReflectionShahid, Kajfez1984}
%\ref{thesisGao}
\begin{eqnarray}
\Gamma_{i} &=& ae^{-2\pi jf\tau}\Gamma_{d}\left[1-\frac{2\kappa e^{i\phi}}{1+\kappa+j2Q_{o}\delta_{L}}\right] 
\label{gammairef}
\end{eqnarray}
with % the shifted resonance frequency $\delta_{L}$
\begin{eqnarray}
\label{sigmaL}
\delta_{L} &=& \delta_{i}-\frac{X_{e}R_{o}}{2Q_{o(}Z_{o}^{2}+X_{e}^{2})} \\
\delta_{i} &=& \frac{\omega-\omega_{o}}{\omega_{o}}
\end{eqnarray}

\begin{figure}[h!]
% figure source: Q_circle_simulation_ResonanceOnly, matlab online folder
\center
\includegraphics[scale=0.4]{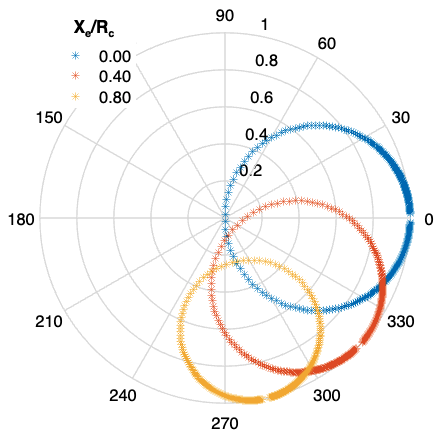}
\caption{\footnotesize{Frequency sweep response of the resonator in phase space. The $S_{11}$ data traced out a circle around the resonance frequency $f_{o}$}. Points outside of the resonance  circle is part of the larger circle that is defined by the off resonance coupling coefficient $\Gamma_{d}$. The coupling reactance $X_{e}$ rotates the resonance circle away from the real axis while shrink the circle diameter at the same time. For $X_{e}=0$ (blue) the center of the circle lies on the real axis. For larger $X_{e}$ (red and yellow) the circle is rotated further away from the real axis clockwise and shrink to a smaller circle due to less coupling of the resonator to the measurement line.}  
\label{QPolarBefore}
\end{figure}

Due to the coupling reactance $X_{e}$ original resonant frequency $\omega_{o}$ is shifted down by $X_{e}R_{o}\omega_{o}/[2Q_{o}(Z_{o}^{2}+X_{e}^{2})]$. $a$  accounts for the overall loss and amplification gain of the line. $\tau$ is the cable delay and it tilts the phase of reflected signal by a slope $2\pi\tau$. Without any cable delay the resonance traces out a circle in the complex plane but the cable delay deforms this circle to a loop like curve. $\phi$ stands for the asymmetry in the resonance due to impedance mismatch. Removing this effect is further elaborated in part \ref{asyres}. $\Gamma_{d}$, defined as in Eq.~\ref{gammaD}, is the decoupled reflection coefficient away from resonance. Together with cable phase delay $\tau$ it traces out a larger circle with wide enough measurement band width. $\kappa$ (Eq.~\ref{kapDef}) is the coupling coefficient which tells the strength the resonator is coupled to the external circuit. Both $\gamma_{d}$ and $\kappa$ are function of $X_{e}$ and figure \ref{QPolarBefore} shows its effect on the resonance circle in phase space. It rotates the resonance circle away from the real axis by angle $\alpha$ as well as shrinks the resonance circle at the same time. Larger $X_{e}$ will rotate the circle further away from the real axis and shrink it to a smaller circle.
\begin{eqnarray}
\Gamma_{d} &=& \left(\frac{jX_{e}-Z_{o}}{jX_{e}+Z_{o}}\right)=e^{-2j~\text{atan}(X_{e}/Z_{o})} = e^{-j\alpha}
\label{gammaD}
\end{eqnarray}
\begin{eqnarray}
%Q_{L} &=& \frac{Q_{o}}{1+\kappa}
\kappa &=& \frac{Q_{o}}{Q_{c}}
\label{kapDef}
\end{eqnarray}
Definition of other parameters are illustrated in figure \ref{cTranRef}
\begin{comment}
\begin{figure}[h!]
% figure source: 'Trans and Ref Resonators.ppt',  juliang.li@anl.gov drive
%\center
   \begin{subfigure}[t]{0.22\textwidth}
   \raisebox{-\height}{\includegraphics[width=\textwidth]{cirRef.png}}    
   \end{subfigure}
   \hfill
   \begin{subfigure}[t]{0.22\textwidth}
   \raisebox{-\height}{\includegraphics[width=\textwidth]{cirRefPara.png}}
   \end{subfigure}
      \begin{subfigure}[t]{0.2\textwidth}
   \raisebox{-\height}{\includegraphics[width=\textwidth]{cirTran.PNG}}    
   \end{subfigure}
   \hfill
   \begin{subfigure}[t]{0.2\textwidth}
   \raisebox{-\height}{\includegraphics[width=\textwidth]{cirTranPara.png}}
   \end{subfigure}
\caption{\footnotesize{schematic of two typical resonator measurement circuits: reflective (\textbf{a}) and transmission (\textbf{b}). The right figures are the corresponding equivalent circuits with definition of the electronic components.}}
\label{cTranRef}
\end{figure}
\end{comment}
%\begin{center}
%Text
%\end{center}
%\vspace*{2ex}  % The spacing above but without the minus.
\section{fitting procedures}
\subsection{Removing cable delay}
\begin{figure}[h!]
\begin{subfigure}[t]{0.2\textwidth}
\begin{overpic}[abs,unit=1pt,scale=.35]{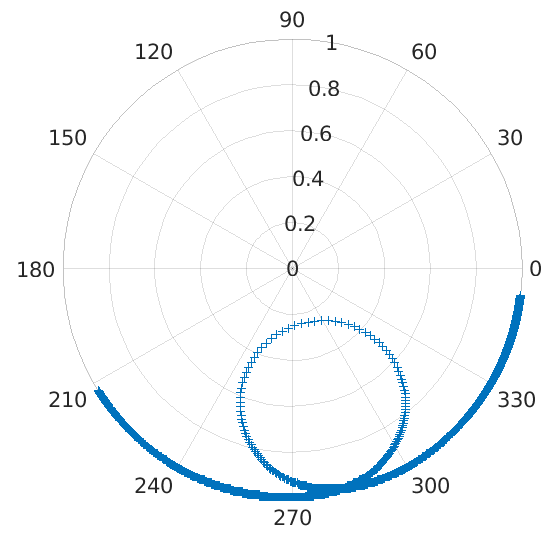}
\put(3,80){\color{black}{(a)}}
\end{overpic}
\end{subfigure}
\hfill
\begin{subfigure}[t]{0.25\textwidth}
\begin{overpic}[abs,unit=1pt,scale=.35]{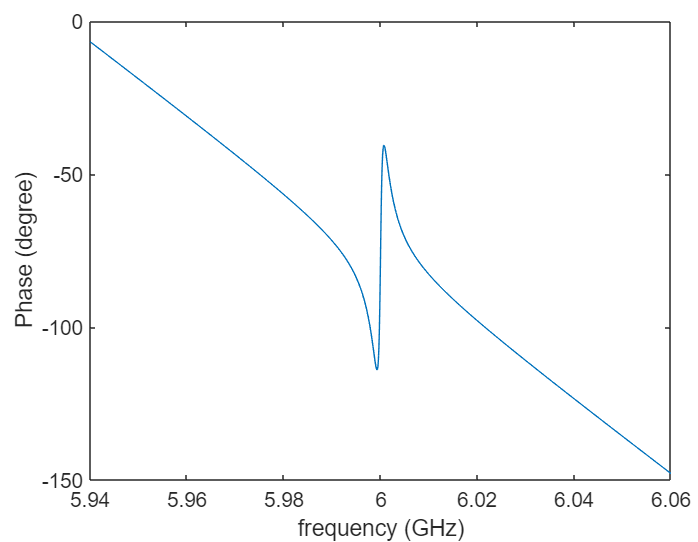}
\put(92,80){\color{black}{(b)}}
\end{overpic}
\end{subfigure}
\caption{\footnotesize{\textbf{(a)} Measurement with electric delay of $\tau=6.64$ nsec. \textbf{(b)} Phase response around the resonance point. The slope of the phase represents  the cable delay and can be used to correct the delay with a linear fit of the phase to the frequency.}}
\label{Cfull}
\end{figure}

\begin{figure}[h!]
\begin{subfigure}[t]{0.2\textwidth}
%   \raisebox{-\height}{\includegraphics[width=\textwidth]{CircleFlat.png}}  
\begin{overpic}[abs,unit=1pt,scale=.35]{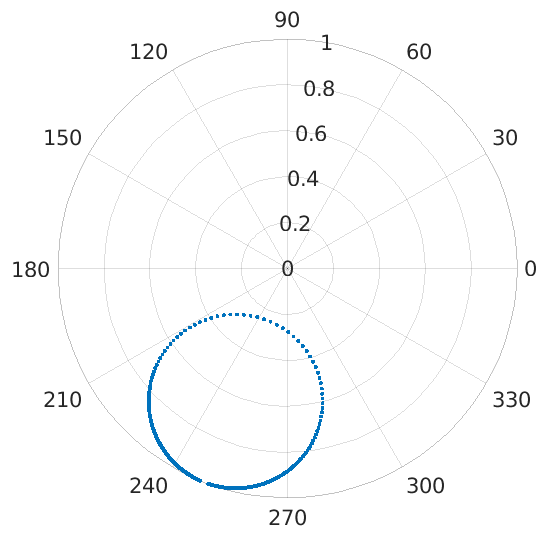}
\put(3,80){\color{black}{(a)}}
\end{overpic}
\end{subfigure}
\hfill
\begin{subfigure}[t]{0.25\textwidth}
%   \raisebox{-\height}{\includegraphics[width=\textwidth]{PhaExpFlat.png}}  
\begin{overpic}[abs,unit=1pt,scale=.35]{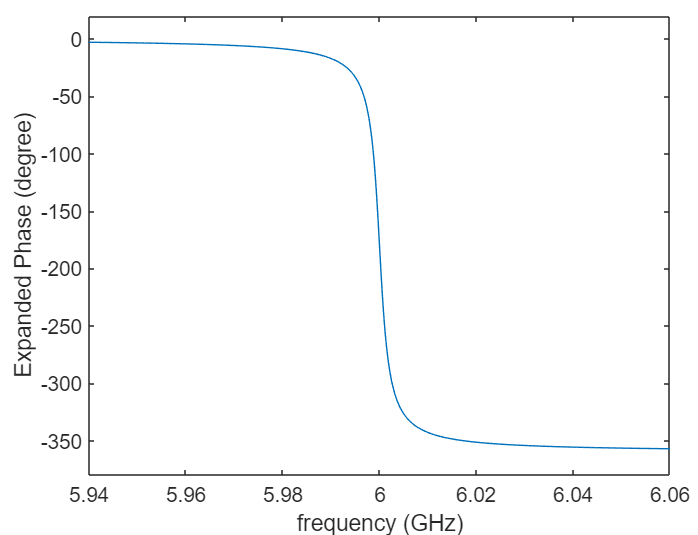}
\put(92,80){\color{black}{(b)}}
\end{overpic}
\end{subfigure}
\caption{\footnotesize{After removing the cable delay the off resonance points shrink to a single point and only the resonance circle is left \textbf{(a)}. The phase away from the resonance is independent of frequency \textbf{(b)}}}
\label{cFlat}
\end{figure}

The cable delay deforms the resonance circle to a loop shape as illustrated in figure \ref{Cfull} (a). In phase format the delay was indicated as a sloped phase away from the resonance point as shown in figure(\ref{Cfull}) (b). The cable delay is equal to the slope divided by $2\pi$ and can be retrieved by linear fit to the line segment either before or after the resonance. The two slopes from either side of the resonance are generally not equal due to impedance mismatch in the microwave line which could be retrieved in step \ref{asyres}. Preliminary solution is taking the average of the two slopes or linear fit the two line segments together. Once the impedance mismatch is determined at step (\ref{asyres}) and corrected a linear fit to the phase plot could be fitted to the mismatched corrected data to tune up the results.
\begin{comment}
\begin{figure}[h!]
\begin{subfigure}[t]{0.2\textwidth}
\raisebox{-\height}{\includegraphics[width=\textwidth]{CircleFull.png}}
\end{subfigure}
\hfill
\begin{subfigure}[t]{0.25\textwidth}
\raisebox{-\height}{\includegraphics[width=\textwidth]{PhaExpFull.png}}
\end{subfigure}
\caption{\footnotesize{\textbf{(a)} Measurement with electric delay of $\tau=6.64$ nsec. \textbf{(b)} Phase response around the resonance point. The slope of the phase represents  the cable delay and can be used to correct the delay with a linear fit of the phase to the frequency.}}
\label{Cfull}
\end{figure}
\end{comment}
After removing cable delay the over all phase is  flat as shown in figure \ref{cFlat} (b) and the cross curves on either side of the resonance frequency in polar display  will shrink to a semicircle as in figure \ref{cFlat} (a).\\
\begin{comment}
\begin{figure}[h!]
\begin{subfigure}[t]{0.2\textwidth}
   \raisebox{-\height}{\includegraphics[width=\textwidth]{CircleFlat.png}}    
\end{subfigure}
\hfill
\begin{subfigure}[t]{0.25\textwidth}
   \raisebox{-\height}{\includegraphics[width=\textwidth]{PhaExpFlat.png}}    
\end{subfigure}
\caption{\footnotesize{After removing the cable delay the off resonance points shrink to a single point and only the resonance circle is left \textbf{(a)}. The phase away from the resonance is independent of frequency \textbf{(b)}}}
\label{cFlat}
\end{figure}
\end{comment}
\subsection{Circle fit}
The resonance circle is rotated away from the real axis with angle $\alpha$ due to the coupling reactance $X_{e}$ (Eq.~\ref{gammaD}). 
A circle is fitted to the data to extract the radius and center of the resonance circle. The resonance circle is shifted to the center of the phase space (figure~\ref{cTrans} \textbf{a}) and (if impedance mismatch is not present) the resonance point is aligned with the real axis (figure~\ref{cTrans} \textbf{b}). Angle $\alpha$ is calculated with Eq.~\ref{alphaC} with $C_x$ and $C_y$ being the coordinates of the fitted circle centers.
\begin{eqnarray}
\label{alphaC}
\alpha &=& \text{atan}\left(\frac{C_y}{C_x}\right)
%d &=& 2R
\end{eqnarray}
\subsection{Translation to the origin}
\begin{figure}[h!]
\begin{subfigure}[t]{0.23\textwidth}
%   \raisebox{-\height}{\includegraphics[width=\textwidth]{CircleShift.png}} 
\begin{overpic}[abs,unit=1pt,scale=.45]{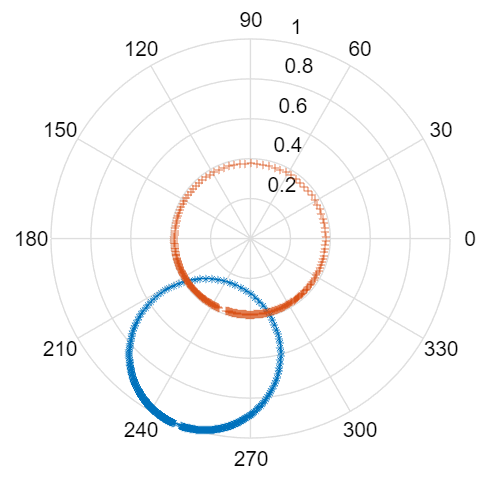}
\put(3,100){\color{black}{(a)}}
\end{overpic}
\end{subfigure}
\hfill
\begin{subfigure}[t]{0.23\textwidth}
%   \raisebox{-\height}{\includegraphics[width=\textwidth]{CircleRotate.png}}
\begin{overpic}[abs,unit=1pt,scale=.45]{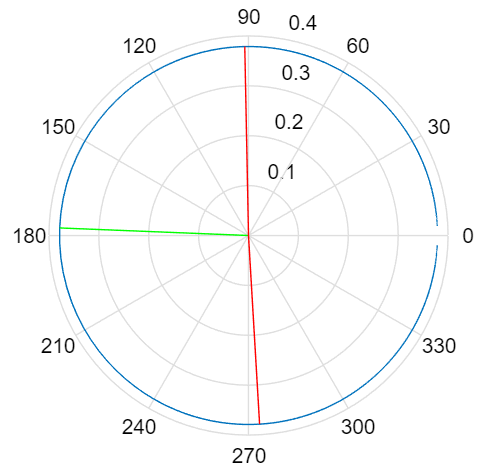}
\put(92,100){\color{black}{(b)}}
\end{overpic}
\end{subfigure}
\caption{\footnotesize{\textbf{a.} The resonance circle (blue color) shifted to the origin (brown color). \textbf{b.} the circle is rotated by angle $-\alpha$ to align with the read axis. The green lines points to the resonance frequency $f_{\text{o}}$ point and the two red lines connects to the $\pm 45^{\circ}$ away from the resonance point that is used to calculate $Q_{L}$.}}
\label{cTrans}
\end{figure}
\begin{figure}[h!]
\center
\includegraphics[scale=0.7]{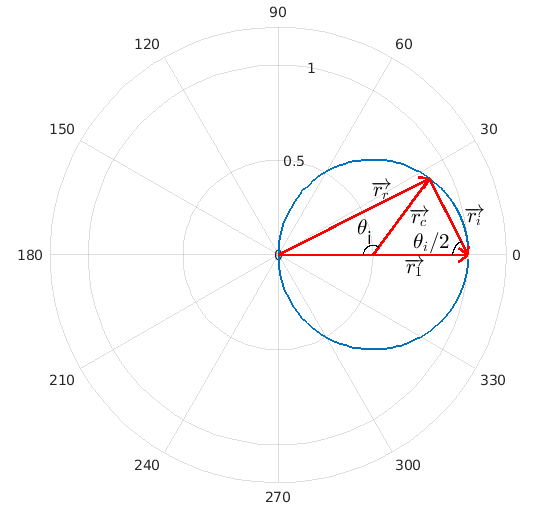}
\caption{\footnotesize{(a) Phase space illustration of vectors in Eq.~\ref{vectorprese}. $\protect\overrightarrow{r_{r}}$ is the vector representing the resonance circle. It could also be represented by vector $\protect\overrightarrow{r_{i}}$ as if the origin of the polar system is at the off resonance point. For the simplicity of calculating the phase the circle will be represented by vector $\protect\overrightarrow{r_{c}}$ by shifting the origin of the phase space to the center of the circle. The phase will be twice of the phase of vector $\protect\overrightarrow{r_{i}}$}}
\label{cirvec}
\end{figure}

\begin{comment}
\begin{figure}[h!]
\begin{subfigure}[t]{0.22\textwidth}
   \raisebox{-\height}{\includegraphics[width=\textwidth]{CircleShift.png}}  
\end{subfigure}
\hfill
\begin{subfigure}[t]{0.22\textwidth}
   \raisebox{-\height}{\includegraphics[width=\textwidth]{CircleRotate.png}}
\end{subfigure}
\caption{\footnotesize{\textbf{a.} The resonance circle (blue color) shifted to the origin (brown color). \textbf{b.} the circle is rotated by angle $-\alpha$ to align with the read axis. The green lines points to the resonance frequency $f_{\text{o}}$ point and the two red lines connects to the $\pm 45^{\circ}$ away from the resonance point that is used to calculate $Q_{L}$.}}
\label{cTrans}
\end{figure}
\end{comment}
After removing the electric delay and correcting the rotation from the coupling reactance.
Eq.~\ref{gammairef} is simplified to
\begin{eqnarray}
\Gamma_{i} &=& a\left[1-\frac{2\kappa}{1+\kappa+j2Q_{o}\delta_{L}}\right] \\
&=& a\left[1-\frac{\frac{2\kappa}{1+\kappa}}{1+j2Q_{L}\delta_{L}}\right]
%\label{gamma_i}
\end{eqnarray}
The equation can also be seen as the sum of vectors in a complex plane
\begin{eqnarray}
\label{vectorprese}
\overrightarrow{r_{r}} &=& \overrightarrow{r_{1}}-\overrightarrow{r_{i}}
\end{eqnarray}
With $\overrightarrow{r_{1}}$ represents the unit vector on the real axis as in figure~\ref{cirvec}, $\overrightarrow{r_{r}}$ is the resonance vector that traces out the resonance circle and $\overrightarrow{r_{i}}$ is the vector corresponding to
\begin{eqnarray}
\Gamma_{i}^{'} &=&a\frac{\frac{2\kappa}{1+\kappa}}{1+j2Q_{L}\delta_{L}}
\end{eqnarray}
which will traces out the same resonance circle as if the off resonance point is the origin of the polar coordinate and the phase of the vector is $\theta_{i}/2=\text{atan}(-j2Q_{L}\delta_{L})$. The same circle will be achieved with vector $\overrightarrow{r_{c}}$ as if the origin of the polar coordinate system is moved to the center of the circle which will has the phase $\theta_{i}$.
When the circle is transferred to the origin and rotated back to align with the real axis, the reflection coefficient is defined by the vector $\overrightarrow{r_{c}}$ which defines the circle in figure \ref{cTrans} \textbf{b}. The diameter $d$ and angle $\theta$ of the circle is related to the resonance circle as:
\begin{eqnarray}
d &=& a\frac{2\kappa}{1+\kappa} \\
\theta &=& 2\text{atan}(-2Q_{L}\delta_{L})
\end{eqnarray}
When one selects two frequencies $f_{3}$ and $f_{4}$ where $\theta_{3,4}=\pm 90^{\circ}$ (figure \ref{cTrans}\textbf{b.}), $Q_{L}$ can be calculated as\cite{Kajfez1984}
\begin{eqnarray}
Q_{L} &=& \frac{f_{L}}{f_{3}-f_{4}} 
\end{eqnarray}
The reflection model (\ref{gammairef}) ignores the coupling loss. When coupling loss present the resonance circle will be distorted and
\begin{eqnarray}
\theta &\approx& 2\text{atan}(-2Q_{L}\delta_{L})
\end{eqnarray}
Least square regression should be applied instead to retrieve $Q_{L}$ with higher accuracy\cite{QReflectionShahid}. After minimizing the quantity:
\begin{eqnarray}
E &=& \sum_{i}^{N}\left[\theta_{i}-2\text{atan}\left(\frac{-2Q_{L}}{f_{L}}(f_{i}-f_{L}\right)\right]^{2}
\label{ephafit}
\end{eqnarray}
$Q_{L}$, $f_{L}$ and $\theta_{o}$ are retrieved together.
\begin{figure}[h]
    \begin{subfigure}[t]{0.22\textwidth}
%        \raisebox{-\height}{\includegraphics[width=\textwidth]{phaseFit.PNG}}
\begin{overpic}[abs,unit=1pt,scale=.265]{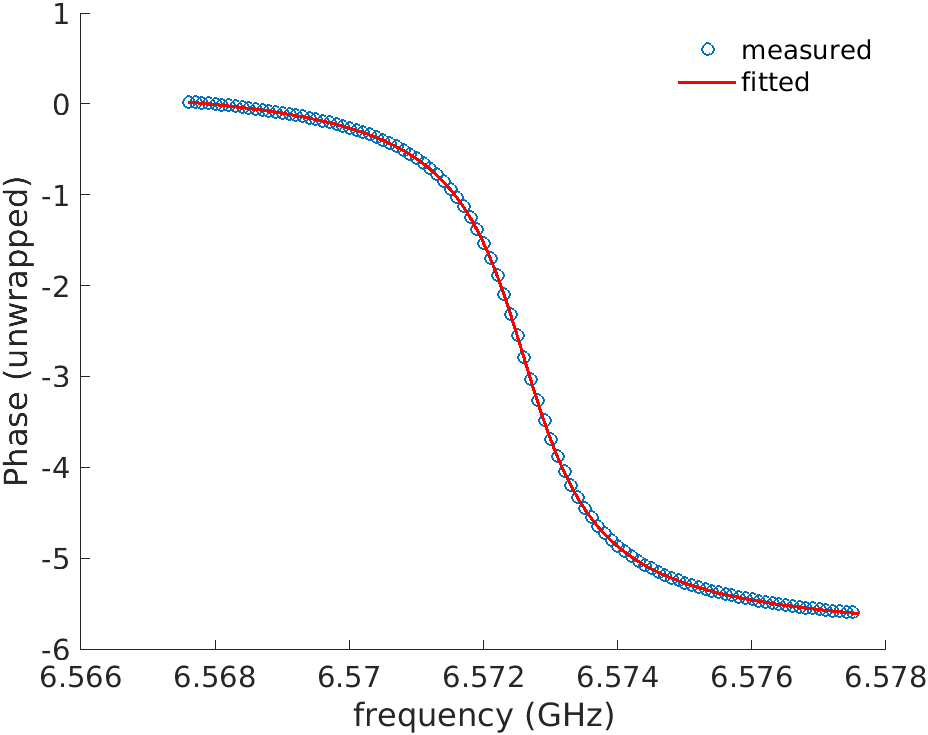}
\put(10,80){\color{black}{(a)}}
\end{overpic}
    \end{subfigure}
    \hfill
    \begin{subfigure}[t]{0.22\textwidth}
%        \raisebox{-\height}{\includegraphics[width=\textwidth]{phaseCom.PNG}}
\begin{overpic}[abs,unit=1pt,scale=.265]{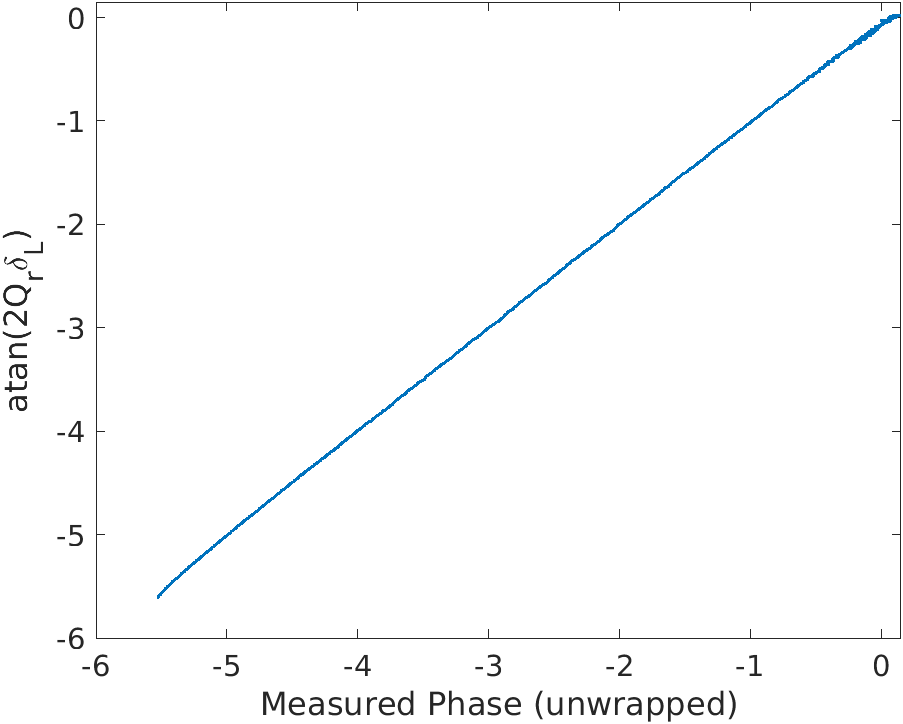}
\put(32,80){\color{black}{(b)}}
\end{overpic}
    \end{subfigure}
    \caption{\textbf{a.} fit of Eq.~\ref{ephafit} for retrieving the $Q_{L}$ and $f_{L}$  \textbf{b.} plot of $\theta_{i}$ vs. $2\text{atan}\left(\frac{-2Q_{L}}{f_{L}}(f_{i}-f_{L}\right)$ with the retrieved $Q_{L}$ and $f_{L}$ to evaluate the fitting quality. A straight line should be achieved with good fitting quality. The upper and lower end of the line is off from the straight line due to impedance mismatch.}
    \label{cphafit}
\end{figure}
\begin{comment}
\begin{figure}[h]
    \begin{subfigure}[t]{0.21\textwidth}
        \raisebox{-\height}{\includegraphics[width=\textwidth]{phaseFit.PNG}}
    \end{subfigure}
    \hfill
    \begin{subfigure}[t]{0.2\textwidth}
        \raisebox{-\height}{\includegraphics[width=\textwidth]{phaseCom.PNG}}
    \end{subfigure}
    \caption{\textbf{a.} fit of Eq.~\ref{ephafit} for retrieving the $Q_{L}$ and $f_{L}$  \textbf{b.} plot of $\theta_{i}$ vs. $2\text{atan}\left(\frac{-2Q_{L}}{f_{L}}(f_{i}-f_{L}\right)$ with the retrieved $Q_{L}$ and $f_{L}$ to evaluate the fitting quality. A straight line should be achieved with good fitting quality. The upper and lower end of the line is off from the straight line due to impedance mismatch.}
    \label{cphafit}
\end{figure}
\end{comment}
\subsection{Resonance asymmetry}
\label{asyres}
\begin{figure}[h]
    \begin{subfigure}[t]{0.195\textwidth}
%        \raisebox{-\height}{\includegraphics[width=\textwidth]{Phim0p4.png}}
\begin{overpic}[abs,unit=1pt,scale=.35]{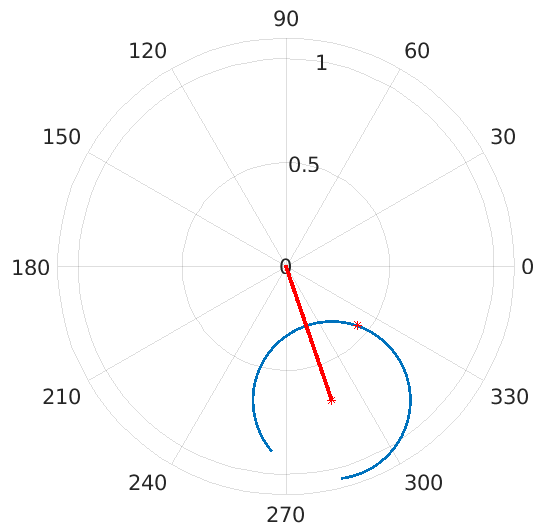}
\put(3,80){\color{black}{(a)}}
\end{overpic}
    \end{subfigure}
    \hfill
    \begin{subfigure}[t]{0.25\textwidth}
%        \raisebox{-\height}{\includegraphics[width=\textwidth]{Phim0p4Amp.png}}
\begin{overpic}[abs,unit=1pt,scale=.35]{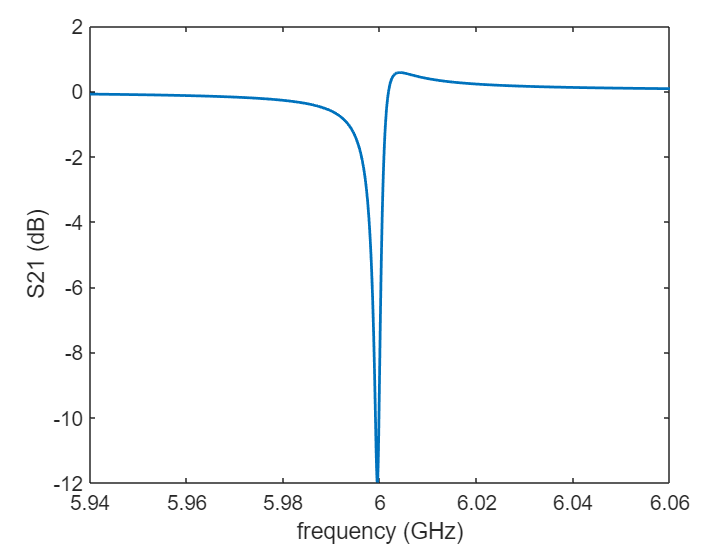}
\put(92,80){\color{black}{(b)}}
\end{overpic}
    \end{subfigure}
    \begin{subfigure}[t]{0.195\textwidth}
%        \raisebox{-\height}{\includegraphics[width=\textwidth]{Phip0p4.png}}
\begin{overpic}[abs,unit=1pt,scale=.35]{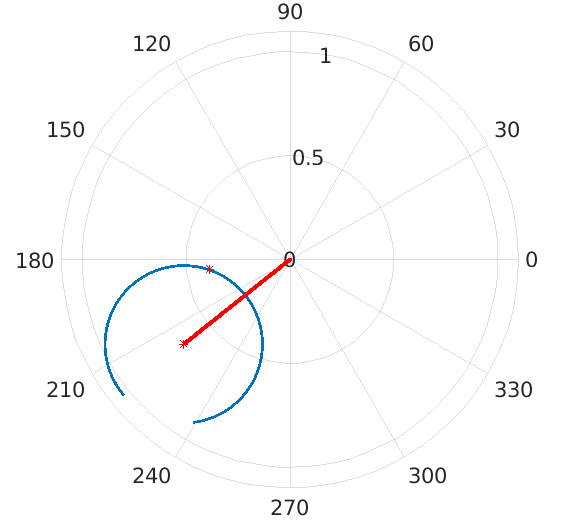}
\put(3,80){\color{black}{(c)}}
\end{overpic}
    \end{subfigure}
    \hfill
    \begin{subfigure}[t]{0.25\textwidth}
%        \raisebox{-\height}{\includegraphics[width=\textwidth]{Phip0p4Amp.png}}
\begin{overpic}[abs,unit=1pt,scale=.35]{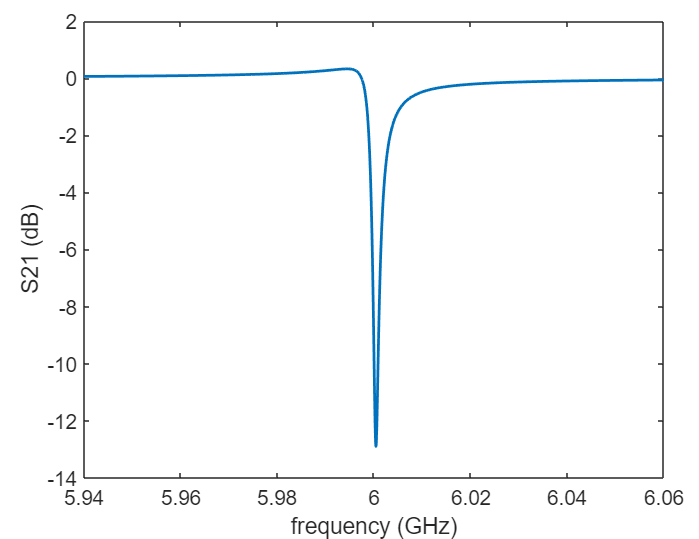}
\put(92,80){\color{black}{(d)}}
\end{overpic}
    \end{subfigure}
    \caption{Illustration of resonance with impedance mismatch. \textbf{a.} polar plot of the resonance circle with $\phi=-0.12\pi$, \textbf{b.} amplitude plot of the resonance dip at the same angle. \textbf{c.} polar plot of the resonance circle with $\phi=-0.12\pi$, \textbf{d.} amplitude plot of the resonance dip at the same angle.}
    \label{circleAsym}
\end{figure}

If impedance mismatch is present in the circuit the resonance will be asymmetric in amplitude plot as shown in figure \ref{circleAsym}. In polar plot the center of the resonance circle is rotated away from the axis connecting the origin of the polar coordinate to the off resonance point by angle $\phi$ (figure \ref{cirq}). To correct this effect the straight forward method is rotating the circle back after recovering the angle $\phi$. This method turns to over estimate the quality factor as the diameter to be used to retrieve the Qs should be the segment along the real axis instead of the diameter of the fitted circle as indicated in figiure \ref{cirq}\cite{khalil2012, probst2015}.
\subsection{Other Qs}
\begin{figure}[h!]
\center
\center
\includegraphics[scale=0.7]{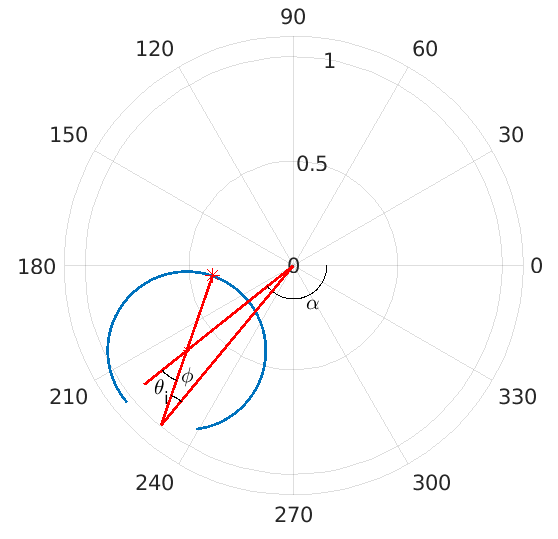}
\caption{\footnotesize{illustration of the three angles $\alpha$, $\phi$ and $\theta_{i}$. $\alpha$ is the rotation of the center of the resonance circle. $\phi$ is the rotation of the circle center away from the segment from the off resonance center to the coordinate origin, which is result of impedance mismatch. With perfect impedance match $\phi=0$ and the circle center align with the off resonance point and the origin. $\theta_{i}$ is the rotation of the resonance point away from the segment between the circle center and the origin.}}
\label{cirq}
\end{figure}

The coupling $Q_{c}$ is calculated through
\begin{eqnarray}
Q_{c} &=& \frac{|z_{c}|+r}{r}Q_{L}
\end{eqnarray}
Here $|z_{c}|+r$ is $a$ in Eq.\ref{gammairef} that accounts for the attenuation and gain in the measurement line. If the circle is rotated due to impedance mismatch, the real $r$ should be the $r'$ illustrated in figure \ref{cirq}. Clearly $r'=r$ if no rotation is present. $r'$ could be calculated through:
\begin{eqnarray}
    2r' &=& 2r\cos\phi
\end{eqnarray}
Finally the internal $Q_{o}$ is calculated through the equation:
\begin{eqnarray}
\frac{1}{Q_{L}} &=& \frac{1}{Q_{c}}+\frac{1}{Q_{o}}
\label{Q3}
\end{eqnarray}
\section{model of transmission measurement}
For transmission or hanger type resonator, the reflected signal measured by the network analyzer is given as: \cite{thesisGao, khalil2012, probst2015, megrant2012}

\begin{eqnarray}
\Gamma_{i} &=& ae^{-2\pi jf\tau}\Gamma_{d}\left[1-\frac{\kappa e^{i\phi}}{1+\kappa+j2Q_{o}\delta_{L}}\right] 
\label{gammaitran}
\end{eqnarray}
\begin{eqnarray}
d &=& \frac{\kappa}{1+\kappa} \\
\tan\theta &=& -2Q_{L}\delta_{L}
\end{eqnarray}
\subsection{Other Qs}
\begin{eqnarray}
Q_{c} &=& \frac{|z_{c}|+r}{2r}Q_{L}
\end{eqnarray}
$Q_{i}$ can be calculated through the same relationship as Eq. \ref{Q3} 
\appendix
\label{appdA}
\section{Calculation of $\omega_{o}$ shift $\Delta \omega$ by $X_{e}$}
introduce $\Delta \omega$ into $\omega_{o}$
\begin{eqnarray}
\delta_{i} &=& \frac{\omega}{\omega_{o}(1+\frac{\Delta\omega}{\omega_{o}})}-1 \\
&=& \frac{\omega}{\omega_{o}}-\frac{\Delta\omega}{\omega_{o}}\frac{\omega}{\omega_{o}}-1 \\
&=& \frac{\omega}{\omega_{o}}-\frac{\Delta\omega}{\omega_{o}}-1 \text{   (for  $\omega=\omega_{o}$ on $2^{nd}$ term)}
\end{eqnarray}
Compare to Eq. \ref{sigmaL}
\begin{eqnarray}
\frac{\Delta\omega}{\omega_{o}} &=& \frac{X_{e}R_{o}}{2Q_{o(}Z_{o}^{2}+X_{e}^{2})}
\end{eqnarray}
\bibliography{PhDThesisRef}
\end{document}